\documentclass[sigconf,nonacm]{acmart}

\usepackage{tabularx}

\renewcommand\footnotetextcopyrightpermission[1]{}
\newcolumntype{L}{>{\raggedright\arraybackslash}X}
\newcolumntype{R}{>{\raggedleft\arraybackslash}X}

\begin{document}

\title{ClawHub Security Signals: When VirusTotal, Static Analysis, and SkillSpector Disagree}
\titlenote{An early, versioned (v1) release. Dataset: \url{https://huggingface.co/datasets/OpenClaw/clawhub-security-signals}.}

\author{Vincent Koc}
\authornote{Corresponding author.}
\affiliation{\institution{OpenClaw Foundation}\country{USA}}
\email{vincent@openclaw.org}

\author{Patrick Erichsen}
\affiliation{\institution{OpenClaw Foundation}\country{USA}}
\email{patrick@openclaw.org}

\author{Jacob Tomlinson}
\affiliation{\institution{NVIDIA}\country{United Kingdom}}
\email{jtomlinson@nvidia.com}

\author{Agustin Rivera}
\affiliation{\institution{NVIDIA}\country{USA}}
\email{arivera@nvidia.com}

\author{Michael Appel}
\affiliation{\institution{NVIDIA}\country{USA}}
\email{mappel@nvidia.com}

\author{Nir Paz}
\affiliation{\institution{NVIDIA}\country{USA}}
\email{npaz@nvidia.com}

\renewcommand{\shortauthors}{Koc et al.}

\begin{abstract}
Agent skills extend AI agents with reusable instructions, tools, scripts, references, and workflows, establishing a security boundary distinct from both model safety and traditional package-malware detection. \textbf{ClawHub Security Signals} is a sanitized dataset of 67,453 latest public OpenClaw skill versions. Each row pairs redacted \texttt{SKILL.md} content and sanitized bundled files where present with a final ClawScan registry verdict and evidence from three scanner families: VirusTotal, static heuristic analysis, and NVIDIA SkillSpector.

Rather than estimating malicious-skill prevalence, we study \textbf{scanner disagreement}. \textbf{The three scanners rarely flag the same skills:} any pair overlaps on at most 10.4\% of their combined positives, only 0.69\% of skills are flagged by all three, and 81.9\% of flagged skills are identified by a single scanner. The disagreement is structured by attack surface. SkillSpector, which raises semantic agentic-risk advisories rather than malware-reputation signals, is positive for 19,209 of 25,504 \texttt{suspicious} rows (75.3\%) but only 14 of 206 \texttt{malicious} rows (6.8\%). The malicious-verdict region shows the inverse profile: 150 of 206 malicious rows (72.8\%) are VirusTotal-positive, consistent with bundled-code malware evidence.

These results show that agent-skill security \textbf{requires layered governance, not single-scanner allow/block decisions}. The corpus is released as a \textbf{sanitized \emph{silver-standard} dataset}: labels are the registry's automated verdicts, not human-annotated ground truth, and the release represents an early, versioned snapshot intended to support the community while a human-annotated subset is developed. Further research is encouraged, including models tailored for skill-security triage.
\end{abstract}


\maketitle

\begin{figure*}[t]
\centering
\includegraphics[width=\textwidth]{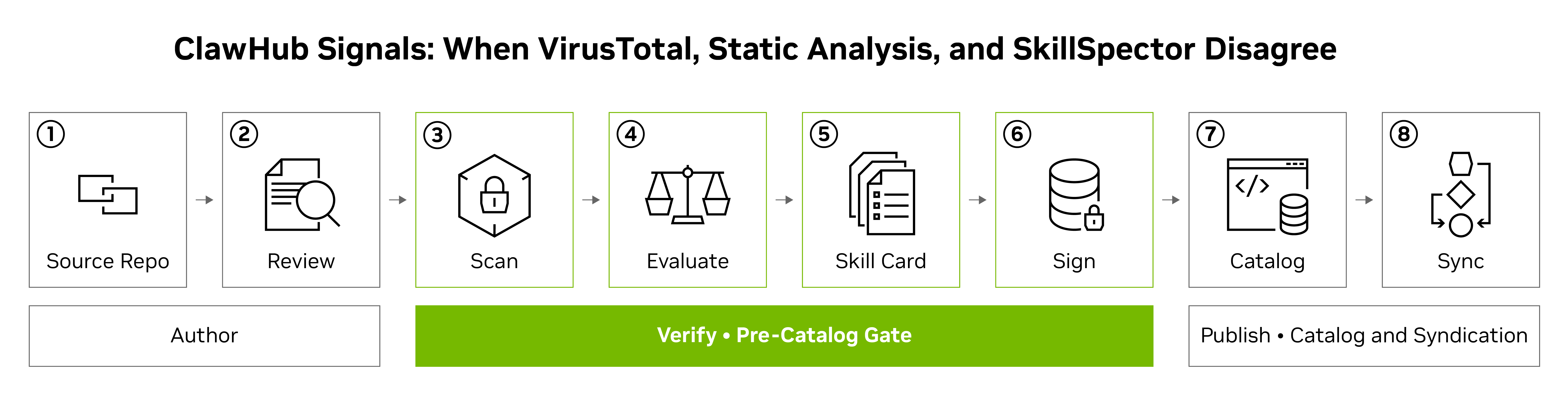}
\caption{ClawHub's skill verification pipeline. The dataset captures ClawScan inputs and verdicts; scanner disagreement is measured among static analysis, VirusTotal, and SkillSpector. Signing is proposed, not yet implemented.}
\label{fig:pipeline}
\Description{Diagram of the ClawHub skill verification pipeline from source repository through review, scanning, ClawScan evaluation, Skill Card generation, proposed signing, catalog publication, and sync.}
\end{figure*}

\section{Introduction}

Agent skills are emerging as a reusable software layer for AI agents, encoding procedural knowledge, tool-use patterns, constraints, dependencies, and, in some cases, executable helper code. Verified agent skills are described by NVIDIA as portable instruction sets that attain trustworthiness only after undergoing scanning, review, signing, and documentation in a skill card~\citep{nvidia2026verifiedskills,nvidia2026trustcontrols}. According to OWASP's Agentic Skills Top 10, skills function as an execution layer that determines what agents do with tools, rather than merely specifying which tools are available~\citep{owasp2026agenticskills}. This characterization positions agent skills as distinct security objects. While a skill may contain benign package content, it can still pose security risks if it grants excessive authority, alters data-flow boundaries, conceals remote-control paths, stores credentials insecurely, or fails to disclose destructive behavior. Conversely, a skill with high agentic risk may remain legitimate and valuable when it is properly documented, signed, and deployed within an appropriate trust context. Consequently, skill trust is not an inherent property of the code itself but is defined by the relationship among the declared purpose, the requested authority, and the agent's operational context.

\paragraph{From prevalence to agreement.} Recent measurement studies have quantified the frequency of vulnerable skills or skills with a malicious registry verdict: empirical analyses have examined tens of thousands of skills for vulnerability patterns~\citep{liu2026agentskills}, behaviorally confirmed malicious samples in a corpus of nearly one hundred thousand skills~\citep{liu2026malicious}, and proposed multi-agent auditing pipelines~\citep{guo2026skillprobe}. These studies establish the widespread nature of the problem. They do not, however, address the central question of this work: when a registry employs multiple independent detectors on the same skill, do these detectors agree, and what are the implications of their disagreement for trust decisions? While a prevalence estimate assumes the reliability of a given detector, this study instead evaluates detectors relative to one another.

\paragraph{Contribution and framing.} We release ClawHub Security Signals, a sanitized snapshot of 67,453 latest public skill versions from the OpenClaw registry, pairing each skill's analyzed bundle content with the registry's final ClawScan verdict and the raw signals from three independent scanner families.\footnote{VirusTotal malware reputation, static analysis, and NVIDIA SkillSpector semantic agentic-risk analysis; see Section~\ref{sec:pipeline}.} We are explicit about epistemics: the verdict is the registry's own automated decision, so we treat the release as a \emph{silver-standard} corpus~\citep{rebholz2010calbc} in which each scanner is a weak-supervision source~\citep{ratner2017snorkel} of unknown accuracy, and we make the lineage of every label explicit rather than presenting it as ground truth. The dataset is a multi-signal \emph{trust} corpus, not a malware corpus.

\paragraph{Disagreement is the finding, not a defect.} We anticipated that three scanners applied to the same 67,453 skills would yield substantial overlap; the actual overlap is minimal (Section~\ref{sec:disagreement}). Any two scanners agree on fewer than one in ten of their combined flags, and only slightly more than chance would predict; only 468 skills (0.69\%) are flagged by all three simultaneously; and 81.9\% of flags originate from a single scanner without corroboration. This does not reflect deficiencies in the scanners themselves. Rather, it demonstrates that different layers of the stack identify distinct risks, and a registry that relies exclusively on any single scanner as the definitive source inherits that scanner's blind spots in their entirety.

\paragraph{The disagreement is structured, not random.} The scanners do not simply diverge; they specialize, and the final verdict reflects which scanner is in scope. Among skills with a \texttt{suspicious} registry verdict, SkillSpector is positive for 75.3\%; among skills with a \texttt{malicious} registry verdict, SkillSpector is positive for 6.8\% and VirusTotal is positive for 72.8\%. Bundled-code malware evidence and semantic agentic-risk evidence are, in this snapshot, different signals that track what each tool inspects: anti-virus engines in VirusTotal, in general, scan all files within a container, whereas SkillSpector reasons about instructions and declared capabilities. Any account that collapses them into one number erases the most useful structure in the data.

\paragraph{An early, living release.} Since the observed disagreement is structural rather than incidental, the logical next step is human adjudication of the disputed cases. Accordingly, v1 is released with automated silver labels, and a future version is planned to include a human-annotated subset that over-samples cases of disagreement (Sections~\ref{sec:adjudication}, \ref{sec:availability}). Early release enables the community to examine the disagreement directly and to develop improved tooling, including models optimized for skill-security triage.

\paragraph{Contributions.}
\begin{itemize}
  \item We release a sanitized, registry-scale silver-standard dataset of 67,453 latest public skill versions with analyzed bundle content, a final verdict, and three-scanner evidence (Section~\ref{sec:dataset}).
  \item We quantify scanner disagreement with raw and chance-corrected agreement, 0.69\% triple-agreement, and 81.9\% single-scanner flags (Section~\ref{sec:disagreement}).
  \item We show that disagreement is structured by attack surface, including a surface-separation result in which malicious-verdict skills are driven by bundled-code malware evidence and are largely outside SkillSpector's semantic agent-risk layer (Section~\ref{sec:disagreement}).
  \item We give a verdict-conditioned analysis of risk categories, signal-magnitude separation, and illustrative cases (Sections~\ref{sec:verdicts}--\ref{sec:cases}), and argue for a layered, systemic defense (Section~\ref{sec:discussion}).
  \item We position the corpus against prior datasets, give an explicit threats-to-validity treatment, and scope a human-adjudicated successor (Sections~\ref{sec:related}, \ref{sec:threats}, \ref{sec:adjudication}).
\end{itemize}

\section{Background and Threat Model}
\label{sec:background}

\subsection{What an agent skill is}

A skill is a portable bundle that tells an agent how to accomplish a task: a \texttt{SKILL.md} document of instructions and triggers, optionally accompanied by helper scripts, reference material, and capability declarations. At install time the bundle becomes part of the agent's effective program. At runtime, if an agent determines that the description of a skill would be useful for the task at hand, it will load the full content of the skill into the context window. Skills can direct the agent to read files, run commands, call APIs, persist state, send messages to external channels, and recover from errors. Because most of a skill is natural language, its risk is often not in a malicious binary but in \emph{what it instructs a capable agent to do} and \emph{how faithfully its prose corresponds to its bundled behavior}.

\subsection{A multifaceted threat model}

Agent-skill risk does not live at one layer, and conflating the layers is a frequent source of confusion. We distinguish three, and note which our scanners actually observe.

\begin{itemize}
  \item \textbf{Artifact layer.} The skill bundle itself: hidden or conflicting instructions, bundled scripts, dangerous shell construction, exposed secrets, untrusted install sources, and mismatch between declared purpose and actual behavior. This is the layer indirect prompt injection targets when a skill ingests external content~\citep{greshake2023indirect,perez2022ignore}.
  \item \textbf{Tool / MCP layer.} The tools, APIs, and Model Context Protocol (MCP) servers a skill expects the agent to use. This layer is about delegated authority: which external systems the skill can reach, what data can flow through them, and whether the agent can trust the tool descriptions it receives. MCP is relevant because servers expose natural-language tool descriptions that agents may treat as instructions, and prior work documents attacks such as tool poisoning and malicious or changed tool descriptions~\citep{invariantlabs2025toolpoisoning,hou2025mcp}. Audits of MCP deployments further show that the server layer can introduce exploitable behavior even when the calling skill is not itself malware~\citep{radosevich2025mcpaudit}.
  \item \textbf{Runtime layer.} What the agent actually does when it executes the skill. Text-level appearance and tool-call behavior can diverge: text safety does not transfer to tool-call safety~\citep{cartagena2026gap}, and confirming runtime behavior generally requires sandboxed execution, benchmarks of agentic attacks and defenses~\citep{zhan2024injecagent,debenedetti2024agentdojo}, and telemetry of tool use~\citep{koc2025telemetry}.
\end{itemize}

Our three scanners work across these layers, and the mapping is the key to the disagreement we report. VirusTotal and static analysis operate at the artifact layer over \emph{bundled code and skills}; SkillSpector reaches into the tool/MCP layer by reasoning about a skill's instructions and declared capabilities; and none fully observes the runtime layer. The disagreement we measure is, in part, three tools sampling different layers of the same object.

\subsection{Why this is a security problem now}

Tool-enabled LLM agents can take high-impact actions: prior work demonstrates agents autonomously exploiting websites and one-day vulnerabilities under experimental conditions~\citep{fang2024hackwebsites,fang2024oneday}, and prompt-injection research shows that instructions embedded in external or retrieved content can manipulate LLM-integrated applications, including tool invocation and data movement~\citep{greshake2023indirect,perez2022ignore,zou2024poisonedrag}. These risks have already appeared in ClawHub itself: Koi Research's ClawHavoc report described an audit of 2,857 ClawHub skills that found 341 malicious skills, later updated to 824 as the marketplace grew, including installer social engineering, obfuscated shell commands, infostealer payloads, reverse shells, and credential exfiltration~\citep{koi2026clawhavoc}. A skill is exactly such a channel: content the agent is expected to trust, follow, and reuse. The path runs from documentation to action, so analysis must account for intent, disclosure, authority, and data movement, not only executable code.

\section{Related Work}
\label{sec:related}

We review five adjacent literatures, then position our corpus against the closest prior datasets. We do not claim a  systematic review; we survey the work that most directly informs the design and interpretation of a multi-scanner skill dataset.

\subsection{Security of agent skills}

The closest prior work measures skill security at scale. \citet{liu2026agentskills} collect tens of thousands of skills and analyze them for vulnerability patterns with a hybrid static-plus-LLM pipeline that is foundational to the semantic scanner (SkillSpector) we rely on; \citet{liu2026malicious} extend this to nearly one hundred thousand skills and behaviorally confirm a set of malicious samples; \citet{guo2026skillprobe} propose a multi-agent auditing system for emerging skill marketplaces; and \citet{li2026securingskills} contribute an architecture and threat taxonomy with concrete configuration-injection cases. These works establish prevalence and detection methods. Our study is complementary: rather than estimating how many skills are vulnerable or malicious under a single detector, we pair a deployed registry's moderation verdict with the raw outputs of three independent scanners and measure their agreement. To our knowledge, this is the first public dataset to expose multi-scanner disagreement on agent skills at registry scale.

\subsection{MCP and tool-layer security}

Because skills route agents toward tools, MCP security is directly relevant. Surveys map the MCP threat landscape, including tool poisoning and ``rug pull'' tool-update attacks~\citep{hou2025mcp}; vendor research documented the first tool-poisoning and tool-description-injection classes~\citep{invariantlabs2025toolpoisoning}; and safety audits show local MCP servers can enable major exploits with client privileges~\citep{radosevich2025mcpaudit}. Standardized benchmarks for indirect prompt injection in tool-using agents~\citep{zhan2024injecagent} and for agent attacks and defenses~\citep{debenedetti2024agentdojo} formalize the runtime-layer risks a skill can trigger.

\subsection{Assistant and LLM extension ecosystems}

Extension marketplaces repeatedly outgrow their trust infrastructure. The Alexa skill-ecosystem study analyzed over 90,000 skills and found weak vetting, arbitrary names, post-approval backend changes, and incomplete permission disclosure~\citep{lentzsch2021alexa}; skill-squatting showed systematic speech-recognition errors could route users to attacker-controlled skills~\citep{kumar2018skillsquatting}. A systematic evaluation of OpenAI's ChatGPT plugin ecosystem raised platform, privacy, and safety concerns rooted in third-party authorship and reliance on natural-language descriptions~\citep{iqbal2024chatgptplugins}. Browser marketplaces show the same pattern: many infringing extensions resemble previously vetted ones and persist after discovery~\citep{moreno2024chromevetting}. Agent skills inherit these dynamics and add executable bundles plus durable, install-time changes to agent behavior.

\subsection{Software supply-chain malware}

Package ecosystems have long been attacked through malicious publication, dependency confusion, typo-squatting, install-time execution, and maintainer compromise. Backstabber's Knife Collection manually analyzed 174 real-world npm, PyPI, and RubyGems packages~\citep{ohm2020backstabber}. A PyPI study found that multi-behavior malicious packages, dominated by information stealing and command execution, were still reachable via mirrors after discovery~\citep{guo2023pypi}; cross-language work showed npm and PyPI malware share install-script, obfuscation, and embedded-URL features~\citep{ladisa2023crosslanguage}. Ecosystem-scale measurement found systemic fragility from transitive dependencies~\citep{zimmermann2019npm}, large-scale measurement established detection and disclosure baselines~\citep{duan2021measuring}, and benchmark efforts argue that malware samples alone are insufficient~\citep{zahan2024malwarebench}. We treat this literature as necessary context but not a sufficient model: a skill's risk can live in natural-language instructions, tool-routing policy, trigger conditions, and purpose/behavior mismatch, not only in bundled code.

\subsection{Scanner disagreement, weak supervision, and trust documentation}

Disagreement among security tools is well documented: a large industrial static-analysis deployment found managing false positives and developer trust as central as detection~\citep{bessey2010billion}, and developer studies found engineers routinely ignore or suppress warnings, limiting any single tool's authority~\citep{johnson2013static}. We extend this from ``tool vs.\ user'' to ``tool vs.\ tool.'' Methodologically, treating multiple noisy detectors as weak-supervision sources to be aggregated rather than trusted individually is the data-programming paradigm~\citep{ratner2017snorkel}, and harmonizing several automatic annotators into a large \emph{silver-standard} corpus, contrasted with a smaller human \emph{gold} standard, is established practice in biomedical NLP~\citep{rebholz2010calbc}. Because our verdict is LLM-produced, we also inherit the known biases and imperfect human agreement of LLM-as-judge setups~\citep{zheng2023judge}. Finally, documentation-first trust has a clear lineage: Datasheets, Data Statements, and Model Cards argue ML artifacts need explicit provenance and risk statements~\citep{gebru2021datasheets,bender2018datastatements,mitchell2019modelcards}; NIST's AI RMF frames trustworthy AI as governance and measurement rather than a binary property~\citep{nist2023airmf}; and Skill Cards apply this lineage to agent capabilities~\citep{nvidia2026skillcards}.

\subsection{Positioning}

Table~\ref{tab:positioning} situates our corpus. Two concurrent agent-skill studies exceed it in raw scale; our distinctive contribution is the combination of a deployed-registry moderation verdict with \emph{multiple} independent scanner signals, released publicly so that disagreement is directly observable.

\begin{table*}[t]
\centering
\caption{Positioning ClawHub Security Signals against prior security datasets for package, extension, and agent-skill ecosystems. ``Signals'' is the security evidence released per item; our differentiator is the public pairing of a registry verdict with multiple independent scanner signals, enabling disagreement analysis.}
\label{tab:positioning}
\begin{tabularx}{\textwidth}{@{}lLlrLc@{}}
\toprule
Study / corpus & Ecosystem & Unit & Scale & Signals & Public \\
\midrule
Ohm et al.~\citep{ohm2020backstabber} & npm/PyPI/RubyGems & package & 174 & manual & yes \\
Zimmermann et al.~\citep{zimmermann2019npm} & npm & package & ecosystem & structural & n/a \\
Lentzsch et al.~\citep{lentzsch2021alexa} & voice skills & skill & 90{,}194 & policy/permission & partial \\
Iqbal et al.~\citep{iqbal2024chatgptplugins} & LLM plugins & plugin & hundreds & framework & no \\
Liu et al.\ (vulnerability)~\citep{liu2026agentskills} & agent skills & skill & 31{,}132 & static+LLM & partial \\
Liu et al.\ (malicious)~\citep{liu2026malicious} & agent skills & skill & 98{,}380 & behavioral & partial \\
\textbf{This work} & \textbf{agent skills} & \textbf{skill} & \textbf{67{,}453} & \textbf{3 scanner signals + verdict} & \textbf{yes} \\
\bottomrule
\end{tabularx}
\end{table*}

\section{The ClawScan Verification Pipeline}
\label{sec:pipeline}

Figure~\ref{fig:pipeline} shows where the dataset's signals are produced. A skill can enter ClawHub as a linked source artifact or as an uploaded bundle through the publisher UI, then pass a pre-catalog verification gate (\textsc{Scan} $\rightarrow$ \textsc{Evaluate} $\rightarrow$ \textsc{Skill Card} $\rightarrow$ \textsc{Sign (proposed)}) before publication in the catalog. At \textsc{Evaluate}, ClawScan consumes the three scanner outputs together with provenance, metadata, and moderation context, and emits a single registry verdict plus a Skill Card. The disagreement we study is the disagreement \emph{among the inputs} to that step.

\paragraph{The three scanner families, and what each inspects.}
The scanners are not noisier or cleaner versions of one another; they look at different things, which is central to our results.
\emph{Static analysis} emits code- and text-pattern findings over the bundle, such as dangerous execution, credential access, exposed secret literals, dynamic code execution, and untrusted install sources.
\emph{VirusTotal} contributes traditional malware and reputation evidence: it aggregates the verdicts of a large set (on the order of seventy) of third-party antivirus engines and URL/domain reputation services over the bundled files, returning per-engine detections and an aggregate detection ratio. It is signature- and reputation-oriented, targeting \emph{bundled executable code}.
In this pipeline, \emph{SkillSpector}~\citep{nvidia2026skillspector,skillspector2026prepublication} contributes semantic agentic-risk analysis over the skill's instructions, declared capabilities, and available skill metadata, producing scored, severity-tagged advisories across categories such as MCP least-privilege, tool poisoning, data ex-filtration, dangerous code execution, rogue-agent behavior, and supply-chain risk. Its hybrid static-plus-LLM methodology builds on foundational large-scale skill-vulnerability analysis~\citep{liu2026agentskills}. SkillSpector findings are advisory risk signals, not accusations and not install-blocking verdicts by themselves. A SkillSpector issue often indicates a meaningful blast radius rather than abuse.

\begin{table*}[!t]
\centering
\caption{Scanner families observe different security surfaces. The columns describe complementary roles in a layered trust pipeline, not competing definitions of maliciousness.}
\label{tab:scanner-surfaces}
\begin{tabular}{>{\raggedright\arraybackslash}p{0.17\textwidth}>{\raggedright\arraybackslash}p{0.22\textwidth}>{\raggedright\arraybackslash}p{0.28\textwidth}>{\raggedright\arraybackslash}p{0.25\textwidth}}
\toprule
Scanner family & Primary surface & Primary Use & Known blind spot \\
\midrule
VirusTotal & Bundled files and reputation & Known malware, suspicious binaries, URL and domain reputation & Agentic intent, over-broad authority, and under-disclosed behavior \\
Static analysis & Bundle text and code patterns & Dangerous APIs, exposed secrets, shell patterns, suspicious install sources & Purpose, disclosure, and user-intent context \\
SkillSpector & Instructions, declared capabilities, and available skill metadata & Semantic agentic risk, blast radius, tool-use risk, disclosure mismatch & Runtime behavior, unseen bundled code, and malware reputation \\
ClawScan context & Provenance, metadata, and moderation context & Registry posture, confidence, and policy context & Not a standalone scanner family \\
\bottomrule
\end{tabular}
\end{table*}

\paragraph{Defining ``positive.''} A scanner is \emph{positive} on a skill when its status is \texttt{suspicious} or \texttt{malicious}; \texttt{clean}, \texttt{stale}, \texttt{error}, and missing statuses are non-positive. This conservative definition is used for every overlap and agreement statistic below. We use ``positive'' rather than ``detection'' deliberately: a positive is evidence to weigh, not a confirmed finding.

\section{Dataset Construction}
\label{sec:dataset}

\subsection{Source, scope, and cleaning}

We constructed the snapshot from \texttt{clawhub.ai} on 31 May 2026. ClawHub, like all public OpenClaw projects at the time of writing, is released under the permissive MIT license, which permits redistribution of the sanitized signals. The source snapshot contains 187,423 public source-artifact rows and 67,478 normalized latest public skill artifacts. The viewer corpus contains 67,453 latest public skill rows with a ClawScan verdict, split deterministically into 47,262 train, 10,076 validation, 6,747 test, and 3,368 evaluation-holdout rows. The 25-row difference reflects normalized artifacts that did not have a complete releasable ClawScan verdict record after validation and were therefore excluded from the viewer corpus. The corpus does include skills that have been independently, human-verified as malicious; we are not releasing the human labels in this version, choosing instead to publish our initial findings promptly.

As part of data cleaning, the public release includes analyzed public skill content and scanner signals, including redacted \texttt{SKILL.md} content and sanitized bundle-file content where present. This matters because the release is not just a \texttt{SKILL.md}-only text corpus: 13,255 rows (19.65\%) include at least one exported bundle file, 6,785 rows (10.06\%) include code files, and the exported bundle files total 58,516 files and 278.9 MB of sanitized content. Our cleaning methodology included a secret-scanning pass with TruffleHog and redaction of secret-like values; we redacted 387 secret-like values, and validation found zero missing ids, splits, content rows, or verdict rows, zero secret-like text rows, and zero TruffleHog-verified secrets after redaction.

\subsection{Label provenance: a silver standard}

The core field \texttt{clawscan\_verdict} takes values \texttt{clean}, \texttt{suspicious}, or \texttt{malicious}, and is produced by the registry's automated review (OpenAI GPT-5.5 high for 99.6\% of rows, with small remainders from GPT-5-mini and GPT-4.1-mini). ClawScan reports high confidence on 87.1\% of rows, medium on 12.5\%, and low on 0.4\%. We treat the verdict as a \emph{silver} label: the registry's own automated decision, useful and operationally meaningful but not human-adjudicated ground truth. As with any LLM-based judgment, it carries known biases and only imperfect agreement with human reviewers~\citep{zheng2023judge}; we return to the implications, including circularity between an LLM scanner and an LLM verdict, in Section~\ref{sec:threats}.

\subsection{Scanner coverage}

All three scanners run on roughly 98\% of rows, but their resolved-status distributions differ sharply (Table~\ref{tab:coverage}). VirusTotal has a resolved clean/suspicious/malicious status for 65,640 rows (97.3\% of the dataset), including 5,225 positive rows (7.75\% of all rows; 8.0\% of resolved VirusTotal rows), with 233 stale rows and 1,580 rows without a result. SkillSpector has a resolved clean/suspicious status for 66,206 rows (98.2\% of the dataset), including 32,856 advisory-positive rows (48.71\% of all rows; 49.6\% of resolved SkillSpector rows) and 33,350 clean rows (49.44\% of all rows; 50.4\% of resolved SkillSpector rows). Each row also carries the redacted \texttt{SKILL.md}, sanitized bundled files where present, the verdict with confidence and model, per-scanner status summaries (VirusTotal counts; static reason codes; SkillSpector score, severity, issue codes, and categories), the nested ClawScan context, and the split name.

\begin{table*}[t]
\centering
\caption{Scanner coverage and positive rates. Positive share is over all 67,453 rows; SkillSpector advisories are risk signals, not maliciousness labels. VirusTotal has resolved clean/suspicious/malicious status for 65,640 rows; among those resolved rows, 8.0\% are positive.}
\label{tab:coverage}
\begin{tabularx}{\textwidth}{@{}LRRRR@{}}
\toprule
Scanner & Rows with source & Source coverage & Positive rows & Positive share \\
\midrule
VirusTotal      & 65{,}873 & 97.66\% & 5{,}225  & 7.75\% \\
Static analysis & 66{,}185 & 98.12\% & 4{,}434  & 6.57\% \\
SkillSpector    & 66{,}222 & 98.18\% & 32{,}856 & 48.71\% \\
\bottomrule
\end{tabularx}
\end{table*}

\section{Scanner Disagreement}
\label{sec:disagreement}

This is the result we most want readers to take away. Of the 67,453 rows, 35,600 (52.8\%) carry at least one positive scanner signal. The striking finding is how little those positives overlap, and how their overlap is structured by what each scanner inspects.

\subsection{Overlap is small, even after chance correction}

Table~\ref{tab:overlap} reports the joint pattern of positives and pairwise agreement. Raw agreement (Jaccard) never exceeds 0.104 for any pair, and chance-corrected agreement (Cohen's $\kappa$) remains ``slight'' on the Landis--Koch scale ($0.045$--$0.082$). Of the 35,600 rows with any positive, 29,153 (81.9\%) are positive on exactly one scanner and only 468 (1.31\% of positive rows; 0.69\% of all rows) on all three. The $\kappa$ values treat stale, error, and missing statuses as non-positive; every pair remains close to zero after chance correction.

A single scanner is therefore a poor allow/block oracle. Any registry that used one as the final authority would inherit that scanner's blind spots.

\begin{table*}[t]
\centering
\caption{How scanner positives co-occur. Left: joint positive patterns over all rows (an upset-style breakdown). Right: pairwise raw (Jaccard) and chance-corrected (Cohen's $\kappa$) agreement. No pair agrees on more than 10.4\% of its combined positives, and chance-corrected agreement is at most $0.082$.}
\label{tab:overlap}
\begin{tabularx}{\textwidth}{@{}Lrr@{\hspace{3em}}Lrr@{}}
\toprule
None (no positive scanner) & 31{,}853 & 47.22\% & VirusTotal $\cap$ Static       & 0.065 & 0.054 \\
VirusTotal only            & 1{,}821  & 2.70\%  & VirusTotal $\cap$ SkillSpector & 0.094 & 0.045 \\
Static only                & 805      & 1.19\%  & Static $\cap$ SkillSpector     & 0.104 & 0.082 \\
SkillSpector only          & 26{,}527 & 39.33\% & & & \\
VirusTotal + Static        & 118      & 0.17\%  & & & \\
VirusTotal + SkillSpector  & 2{,}818  & 4.18\%  & & & \\
Static + SkillSpector      & 3{,}043  & 4.51\%  & & & \\
All three                  & 468      & 0.69\%  & & & \\
\bottomrule
\end{tabularx}
\end{table*}

\subsection{Disagreement is structured by attack surface}

The scanners do not diverge at random. Table~\ref{tab:withinverdict} cross-tabulates each scanner's positivity against the final verdict. SkillSpector is the dominant positive source in the review-needed region: it raises advisories for 75.3\% of \texttt{suspicious} skills and is the \emph{only} positive scanner on 56.3\% of suspicious skills. The pattern inverts for \texttt{malicious}: VirusTotal flags 72.8\% of skills with a malicious registry verdict while SkillSpector raises advisories for only 6.8\%. This inversion is exactly what the scanner-family roles in Table~\ref{tab:scanner-surfaces} predict: malware lives in bundled code that anti-virus engines used by VirusTotal scan, whereas disclosure and authority risk are determined by SkillSpector's LLM stage.

\begin{table*}[t]
\centering
\caption{Scanner positivity conditioned on the final verdict (percentages within-verdict). ``No positive'' means no scanner reached \texttt{suspicious}/\texttt{malicious} status; for malicious-verdict rows this reflects verdicts driven by provenance and moderation context rather than scanners. The dominant scanner inverts between the review-needed and malicious-verdict regions.}
\label{tab:withinverdict}
\begin{tabularx}{\textwidth}{@{}LrRRRR@{}}
\toprule
Verdict & $n$ & VirusTotal+ & Static+ & SkillSpector+ & No positive \\
\midrule
clean      & 41{,}743 (61.9\%) & 1{,}847 (4.4\%) & 1{,}355 (3.2\%) & 13{,}633 (32.7\%) & 26{,}470 (63.4\%) \\
suspicious & 25{,}504 (37.8\%) & 3{,}228 (12.7\%) & 3{,}053 (12.0\%) & 19{,}209 (75.3\%) & 5{,}333 (20.9\%) \\
malicious  & 206 (0.3\%)       & 150 (72.8\%)    & 26 (12.6\%)     & 14 (6.8\%)        & 50 (24.3\%) \\

\bottomrule
\end{tabularx}
\end{table*}

\subsection{The malicious paradox}

Two facts in the malicious-verdict row of Table~\ref{tab:withinverdict} deserve emphasis. First, SkillSpector's semantic agentic-risk layer is mostly silent on malicious-verdict cases, presumably driven by bundled-code or provenance evidence: the mean SkillSpector issue count for skills with a malicious registry verdict is 0.57 and the median is 0, because 192 of 206 malicious-verdict rows carry no SkillSpector issues. Second, 24.3\% of malicious verdicts have \emph{no} positive scanner signal of any kind; ClawScan reached \texttt{malicious} from provenance, metadata, and moderation context. Both facts follow from where the relevant evidence resides: in bundled executable code, package provenance, and registry moderation context that a \texttt{SKILL.md}-and-capability scanner does not fully observe even when sanitized bundle content is present in the release. The tooling that catches a credential-stealer is not the tooling that catches an over-privileged, under-disclosed automation skill. This is the strongest single argument in the dataset for layered skill governance.

\subsection{Signal magnitude separates the verdicts}

Although SkillSpector positivity is not a final verdict, its reported score separates 2 (clean and suspicious) of the 3 classes, while most malicious-verdict rows fall outside its resolved semantic-advisory surface. Mean SkillSpector score rises from 22.1 (clean) to 59.3 (suspicious), and mean issue count from 1.9 to 6.5; redacted \texttt{SKILL.md} length also grows modestly with risk (median 3{,}955 characters for clean vs.\ 5{,}562 for suspicious). The malicious-verdict class is the exception that proves the point: for the 6.8\% of malicious rows where SkillSpector is positive, the score averages 82.4, while its near-zero issue count reflects the 192 with none. These separations suggest the \emph{magnitude} of the score is informative for the clean/suspicious boundary that constitutes most of the dataset, and is a natural target for a learned triage model.

\section{Verdict Structure and Risk Categories}
\label{sec:verdicts}

\subsection{Verdicts and trust interpretation}

Table~\ref{tab:withinverdict} reports a deliberately non-binary verdict distribution: 61.9\% clean, 37.8\% suspicious, and 0.3\% malicious. The \texttt{suspicious} class is a \emph{review-before-trusting} posture, not an abuse label. It includes skills with unclear disclosure, over-broad authority, scanner disagreement, risky defaults, or a wide blast radius. Three numbers tie disagreement back to trust. First, 32.7\% of clean skills still carry a SkillSpector advisory; this does not contradict the clean registry verdict; it means the skill has risk-relevant properties that may be acceptable when disclosed, purpose-aligned, and bounded by user expectations. Second, 77.2\% of suspicious skills have no static or VirusTotal positive, so the suspicious class is heavily driven by semantic, capability, and disclosure context. Third, 74.8\% of skills with a malicious registry verdict do have a static or VirusTotal positive finding, the region where scanners corroborate one another.

\subsection{What the categories say, and how they shift by verdict}

The most common SkillSpector categories are not classic malware indicators (Table~\ref{tab:categories}); they describe authority, scope, tool semantics, and disclosure. Their \emph{composition} also shifts with the verdict. Data Exfiltration is slightly more common among clean skills (1,196) than suspicious ones (996), because disclosed, purpose-aligned data flow, such as an email summarizer delivering to a configured channel, is legitimate. Dangerous Code Execution shows the opposite skew, concentrating in suspicious skills (1,327 vs.\ 302), as does Tool Poisoning (3,462 vs.\ 1,621). The categories that move a skill toward \texttt{suspicious} are about unsafe \emph{execution and tool manipulation}, not about whether the skill touches sensitive data at all.

\begin{table}[t]
\centering
\caption{Most common SkillSpector categories overall and split across the clean and suspicious verdicts. Counts are row-level category occurrences; a skill may have multiple categories, and columns need not sum to the total because of the small malicious set and null-result rows.}
\label{tab:categories}
\begin{tabularx}{\columnwidth}{@{}Lrrr@{}}
\toprule
Category & Total & Clean & Suspicious \\
\midrule
MCP Least Privilege      & 9{,}641 & 4{,}593 & 5{,}047 \\
MCP Tool Poisoning       & 5{,}084 & 1{,}621 & 3{,}462 \\
Data Exfiltration        & 2{,}192 & 1{,}196 & 996 \\
Dangerous Code Execution & 1{,}629 & 302     & 1{,}327 \\
Rogue Agent              & 1{,}428 & 536     & 891 \\
Supply Chain             & 1{,}336 & 592     & 744 \\
Data Flow                & 976     & 331     & 645 \\
Privilege Escalation     & 792     & 259     & 533 \\
Tool Misuse              & 647     & 231     & 415 \\
Excessive Agency         & 511     & 205     & 306 \\
\bottomrule
\end{tabularx}
\end{table}

\subsection{Static findings and a coarse risk-theme lens}

Static findings are rarer but sharper. The most common reason codes are dangerous execution (1,428 rows), environment-credential access (1,298), exposed secret literals (1,219), dynamic code execution (451), prompt-injection instructions (433), untrusted install sources (250), destructive delete commands (201), potential exfiltration (181), insecure TLS verification (166), and secret exposure via command arguments (121). A small number escalate to malicious-tier static codes, including crypto-mining (29) and stealth-browser abuse (10). A coarse, recall-oriented keyword lens over the redacted skill text shows how pervasive capability-bearing language is: roughly four in five skills (79.8\%) mention sensitive-data or exfiltration-adjacent operations, about a quarter mention persistence or scheduled execution (29.9\%), supply-chain or dependency operations (26.4\%), and network or remote control (25.6\%), and one in five mention overbroad privilege (22.2\%) or insecure secret handling (21.3\%). This lens conveys prevalence of capability-bearing language, not per-skill risk.

\section{Illustrative Cases}
\label{sec:cases}

These examples are illustrative rather than a formal qualitative analysis. Aggregate statistics understate how context-dependent these judgments are. We summarize representative public skills (slugs as published; rationales paraphrased and redacted). The cases also separate malware from moderation. A skill can be policy-blocked because it enables abuse, evasion, or under-disclosed control even when the person installing it is not the direct victim. This is analogous to the potentially unwanted application (PUA) grey zone in endpoint security: Microsoft explicitly separates PUAs from malware while still classifying categories such as evasion software as policy-relevant security signals~\citep{microsoft2026pua}.

\begin{itemize}
  \item \textbf{Clean, high agentic risk.} \texttt{scald/granola} (clean, SkillSpector score 100) transparently syncs meeting notes to local files using the user's existing desktop session token. \texttt{4xiomdev/whoop-central} (clean, score 100) is a coherent health-data integration that nonetheless handles sensitive biometric data and OAuth tokens. Both are correctly clean and correctly carry strong advisories: the advisory describes what the user is accepting, not wrongdoing.
  \item \textbf{Trusted-authors.} \texttt{gumadeiras/roku} (suspicious) is a Roku controller published by a known OpenClaw maintainer. It is genuine and purpose-aligned, yet it bundles under-disclosed Telegram and local-pipe control paths that can issue commands without clear access control. A suspicious flag on a legitimate maintainer's skill is itself an indicator of how hard scanning is: disclosure mismatch, not malice, drives the signal, and trusted authors can still ship control paths that are genuine in intent.
  \item \textbf{Policy-blocked abuse tooling.} \texttt{pkiv/browse} (malicious) openly supports browser automation but explicitly promotes bypassing CAPTCHAs, Cloudflare, and bot detection using stealth browsers, residential proxies, and persistent sessions. This need not mean the installer is the immediate victim. It is closer to hacktool or PUA-style moderation: the artifact is designed to enable unwanted or abusive behavior, so a registry can reasonably refuse distribution even when classic malware scanners are silent.
  \item \textbf{Conflict.} \path{oliveskin/agent-tinman} carries a VirusTotal detection and prompt-injection indicators (``ignore previous instructions'') yet remains \texttt{suspicious} pending human review, illustrating how a malware hit and a final verdict can legitimately diverge.
\end{itemize}

These cases demonstrate that skill trust has multiple facets: malware reputation, static code risk, semantic agentic risk, disclosure, and registry posture can diverge, so summary verdicts should be interpreted with the underlying evidence rather than as standalone ground truth.

\section{OWASP-Aligned Risk Lens}
\label{sec:owasp}

OWASP's GenAI Security Project separates risks for LLM apps, agentic apps, and skills~\citep{owasp2025llm,owasp2026agenticapps,owasp2026agenticskills}. We use these categories as a shared vocabulary for grouping observable evidence (Table~\ref{tab:owaspmap}); we do not claim that any dataset category is an official OWASP label.

\begin{table*}[t]
\centering
\caption{OWASP-aligned risk lens used for analysis. These are grouping labels for observable dataset evidence, not official OWASP labels assigned to individual skills.}
\label{tab:owaspmap}
\small
\begin{tabularx}{\textwidth}{@{}>{\raggedright\arraybackslash}p{\columnwidth}@{\hspace{\columnsep}}L@{}}
\toprule
Risk lens & Dataset evidence grouped under the lens \\
\midrule
Goal hijacking / tool manipulation & MCP Tool Poisoning; Prompt Injection; Trigger Abuse \\
Over-privilege / excessive agency & MCP Least Privilege; Privilege Escalation; Tool Misuse; Excessive Agency \\
Sensitive-data exposure & Data Exfiltration; Data Flow; System Prompt Leakage \\
Unsafe execution & Dangerous Code Execution; static execution findings \\
Supply-chain risk & Supply Chain; untrusted-source install; dependency-not-found findings \\
Persistent trust-state risk & Memory Poisoning; Rogue Agent \\
\bottomrule
\end{tabularx}
\end{table*}

\section{Toward Human Adjudication}
\label{sec:adjudication}

The disagreement in Section~\ref{sec:disagreement} is exactly why automated labels alone cannot close this problem. When three scanners flag largely disjoint sets of skills, 56.3\% of review-needed skills rest on a single semantic agent-risk signal, and 24.3\% of malicious registry verdicts rest on no scanner at all, the appropriate trust posture for a disputed skill is genuinely uncertain and frequently demands human judgment about disclosure, intent, and agentic risk.

We see human adjudication of these disputed cases as the natural next direction. A subsequent version could add a human-annotated subset that over-samples the hard cases this snapshot exposes, single-scanner positives, scanner conflicts, clean-but-advised skills, high-agentic-risk categories, and rows with exported code-bearing bundle files. Rather than forcing one opaque label per skill, such adjudication would record separate dimensions, declared purpose, observed or inferable behavior, privilege and exposure level, external data sinks, secret handling, persistence, hidden-instruction evidence, bundled-code behavior, and MCP/tool interaction risk, from which a final registry posture could be derived. Methodologically, this treats the scanners as weak-supervision sources and the human subset as the instrument that calibrates and bounds the silver labels' error~\citep{ratner2017snorkel,rebholz2010calbc}; because inter-annotator agreement is itself a research object in subjective security labeling~\citep{artstein2008interrater}, we would report annotator disagreement as a first-class result. We describe this as a direction rather than a commitment.

\section{Threats to Validity}
\label{sec:threats}

We follow measurement-study practice and state threats explicitly.

\paragraph{Label provenance.} Verdicts are silver labels from the registry's automated review (OpenAI GPT-5.5 high for 99.6\% of rows). They are not human ground truth, and a different model or moderation configuration could move the clean/suspicious boundary that dominates the data.

\paragraph{Construct validity.} A positive scanner status is evidence, not a confirmed vulnerability; we measure \emph{agreement among detectors}, not \emph{correctness}. Statements about disagreement are robust to verdict error in a way that prevalence claims would not be, and we deliberately avoid the latter.

\paragraph{Circularity.} SkillSpector is partly LLM-based, and the ClawScan verdict is LLM-produced. Correlation between a SkillSpector advisory and the verdict may reflect shared mechanism rather than independent confirmation, and LLM judges carry positional, verbosity, and self-enhancement biases~\citep{zheng2023judge}. This is a reason to study disagreement, which is not inflated by shared mechanism, rather than agreement-with-the-verdict.

\paragraph{Sanitization and reproducibility.} The release is a sanitized research corpus, not a registry mirror. It includes redacted \texttt{SKILL.md} content and sanitized exported bundle files where present, but secret-like values, private identifiers, and private artifact content are removed or redacted. Byte-for-byte reproduction of every scanner decision may therefore require the scanner metadata released with the dataset rather than only the redacted content.

\paragraph{Coverage and agreement bias.} VirusTotal is resolved for 97.3\% of rows in this snapshot, with 233 stale rows and 1,580 rows without a result. ``Not positive'' is still not a human-confirmed clean label, but the earlier large pending-queue caveat no longer drives the agreement statistics; all scanner pairs remain near-zero after chance correction.

\paragraph{Selection and temporal validity.} The corpus is one registry, latest-version only, public skills only, English-heavy, and a single dated snapshot; both skills and scanner versions drift, and VirusTotal scanning is asynchronous.

\section{Data Availability, Licensing, and Maintenance}
\label{sec:availability}

The dataset is released on the Hugging Face Hub, signals-first, with a Gebru-style datasheet~\citep{gebru2021datasheets} and machine-readable metadata documenting composition, collection, intended use, and limitations. ClawHub and all public OpenClaw projects are released under the permissive MIT license at the time of publishing, which covers the sanitized signals and analyzed public skill content we redistribute. We treat the corpus as a \emph{living dataset} in the sense of living systematic reviews~\citep{elliott2014living}: v1 ships automated silver labels, redacted \texttt{SKILL.md} content, and sanitized analyzed bundle content, including code-bearing bundle files where exported; a successor could add a human-annotated subset (Section~\ref{sec:adjudication}), with versioned releases and a changelog. Future releases should also preserve source-artifact revisions, scanner versions or commit identifiers, scanner run timestamps, and model or policy versions where applicable. Deterministic splits and analysis scripts are released for reproducibility. The intended use is scanner, trust, and moderation research; we ask consumers not to treat silver labels as ground truth in downstream claims.

\section{Discussion}
\label{sec:discussion}

\paragraph{A layered, systemic defense is essential.} The observed disagreement and its underlying structure indicate a clear design principle: since each scanner examines a distinct attack surface, no single component can comprehensively secure agent skills. Effective defense should integrate complementary components, each mapped to specific surfaces. Reputation and signature scanning are most effective for detecting bundled-code malware; static analysis addresses code-pattern risks; capability-aware analysis targets semantic authority and disclosure risks; and runtime behavior, which is not fully observed by any current scanner, requires sandboxed execution and telemetry of agent tool use~\citep{debenedetti2024agentdojo,koc2025telemetry}. Functional, behavioral testing of a skill in a sandbox, executing it and monitoring tool calls and data movement, provides the most accurate signal for risk assessment, though it is resource-intensive and challenging to implement at registry scale. Therefore, it should be considered a valuable, albeit costly, complement to the more scalable static and semantic signals analyzed in this study. The empirical disagreement reported here supports the view that skill security is fundamentally a systems problem, best addressed by a layered pipeline that integrates multiple components and transparently presents the evidence underlying each verdict, rather than relying on a single allow/block mechanism.

\paragraph{Advisories are not accusations, and \texttt{suspicious} is not \texttt{malicious}.} One-third of clean skills carry an advisory, and most suspicious skills have no static or VirusTotal positive. Collapsing ``has an advisory'' into ``is bad,'' or ``suspicious'' into ``malicious,'' would discard the most useful structure in the data. The trust question for an advised skill is whether its capabilities are disclosed, purpose-aligned, least-privileged, and bounded by clear user expectations.

\paragraph{An opportunity for skill-security triage models.} Because the disagreement is structured and large-scale, it is a natural target for specialized models: triaging skill risk from sanitized bundle content plus scanner metadata, predicting when a semantic advisory should trigger review, require documentation, contribute to registry posture, or draft a Skill Card summary. The score separations in Section~\ref{sec:disagreement} indicate such models have real signal to learn from, and the weak-supervision framing~\citep{ratner2017snorkel} suggests aggregating the disagreeing scanners into a denoised label model as a concrete first baseline.

\section{Ethics and Responsible Disclosure}

The dataset includes sanitized analyzed skill content and is intended for scanner, trust, and moderation research, not offensive use or exploit reproduction. Case-study slugs are public registry identifiers; their rationales are paraphrased and redacted, and we name them only to illustrate trust categories, not to attribute wrongdoing, especially for clean and suspicious skills, where a positive signal is explicitly not an accusation. Skills with malicious registry verdicts are handled through the registry's existing moderation and takedown process; we do not publish exploit-enabling detail. Researchers should avoid deanonymizing publishers beyond the public slugs already present in the registry and should not use the data to target skill authors. Because labels are silver-standard, downstream users should not present them as adjudicated maliciousness.

\section{Conclusion}

Agent skills bring a familiar malware and potentially unwanted application (PUA) detection problem into the agent setting, with evidence distributed across prose instructions, configuration, tool wiring, and executable code rather than concentrated in a binary. Most skills are benign; a small fraction are clearly malicious; and a consequential middle ground is context-dependent, where identical capabilities may be legitimate or unacceptable depending on authorship, disclosure, and the authority granted to the agent. In this setting, trust is commonly established through review, which is similar to other package repositories, such as PyPI, that must moderate malicious packages and sometimes remove them~\citep{guo2023pypi}. The central finding of this work is that the three scanners feeding the registry rarely agree on which skills warrant a positive signal, and their disagreement is structured by attack surface rather than noise: the dominant scanner inverts between the review-needed and malicious-verdict regions along the boundary of what each tool actually inspects. We release the snapshot as an early silver-standard measurement, not a human-adjudicated corpus, because that distinction matters for any downstream use of the labels. Trustworthy skill ecosystems need transparent Skill Cards, multi-signal scanner evidence, provenance, signing, and governance that separates potential risk from the final verdict, with human adjudication for the disputed middle. We release this dataset to help the community study the disagreement and build the layered, systemic tooling, including tuned skill-security models, that the problem now demands.

\begin{acks}
We thank the security and open-source research communities, whose open work this dataset builds on; the OpenClaw Foundation and NVIDIA teams who built and operate the ClawHub verification pipeline; and, above all, the many contributors who have created and published skills on ClawHub, whose work makes a study like this possible.
\end{acks}

\bibliographystyle{ACM-Reference-Format}
\bibliography{references}

\end{document}